\documentclass[aip,reprint]{revtex4-1}

\usepackage[pdftex]{graphics}
\usepackage{amssymb,amsmath}

\begin{document}

\title{Fast quasi-centroid molecular dynamics}
\author{Theo Fletcher}
\email{theo.fletcher@chem.ox.ac.uk}
\affiliation{Department of Chemistry, University of Oxford, Physical and Theoretical Chemistry Laboratory, South Parks Road, Oxford, OX1 3QZ, UK}
\author{Andrew Zhu}
\affiliation{Department of Chemistry, University of Oxford, Physical and Theoretical Chemistry Laboratory, South Parks Road, Oxford, OX1 3QZ, UK}
\author{Joseph E. Lawrence}
\affiliation{Laboratory of Physical Chemistry, ETH Z\"urich, 8093 Z\"urich, Switzerland}
\author{David E. Manolopoulos}
\affiliation{Department of Chemistry, University of Oxford, Physical and Theoretical Chemistry Laboratory, South Parks Road, Oxford, OX1 3QZ, UK}
\begin{abstract}
We describe a fast implementation of the quasi-centroid molecular dynamics (QCMD) method in which the quasi-centroid potential of mean force is approximated as a separable correction to the classical interaction potential. This correction is obtained by first calculating quasi-centroid radial and angular distribution functions in a short path integral molecular dynamics simulation, and then using iterative Boltzmann inversion to obtain an effective classical potential that reproduces these distribution functions in a classical NVT simulation. We illustrate this approach with example applications to the vibrational spectra of gas phase molecules, obtaining excellent agreement with QCMD reference calculations for water and ammonia and good agreement with the quantum mechanical vibrational spectrum of methane. \end{abstract}

\maketitle

\section{Introduction}

There is considerable current interest in developing more accurate and efficient imaginary time path integral methods for simulating vibrational spectra. The reason for this interest is that neither of the standard methods for including nuclear quantum effects in simulations of dynamical properties -- centroid molecular dynamics\cite{cao-voth94bjcp,cao-voth94djcp} (CMD) or ring polymer molecular dynamics\cite{crai-mano04jcp,habe+13arpc} (RPMD) -- is good enough for this purpose. While these methods are useful for calculating zero-frequency observables such as diffusion coefficients and chemical reaction rate coefficients, they both have serious issues when applied to the calculation of vibrational spectra. These issues were first identified and clarified by Marx and co-workers,\cite{witt+09jcp,ivan+10jcp} who showed that CMD suffers from a curvature problem that causes the frequencies of stretching bands to red shift towards zero with decreasing temperature, and that RPMD suffers from spurious resonances associated with the internal modes of the ring polymer.

It has since been shown that the spurious resonances can be eliminated from RPMD by attaching a thermostat to the internal modes, as is done in the thermostatted ring polymer molecular dynamics (TRPMD) method.\cite{ross+14jcp} However, the damping that this introduces causes an undesirable broadening of spectral features that becomes increasingly pronounced at lower temperatures.\cite{ross+14jcp}  A more promising solution is offered by the recent development of quasi-centroid molecular dynamics (QCMD),\cite{tren+19jcp} in which the Cartesian centroids of ordinary CMD are replaced by curvilinear centroids. Althorpe and co-workers have shown that this eliminates the curvature problem of CMD and gives vibrational spectra in remarkably good agreement with quantum mechanical reference calculations, in terms of both the frequencies and the line shapes of the vibrational bands.\cite{tren+19jcp,bens+20fd,hagg+21jcp} So much so that this is undoubtedly the most accurate RPMD or CMD-based method that has yet been suggested for calculating vibrational spectra.

The numerical implementation of QCMD is, however, rather expensive. For applications to systems with more than a few degrees of freedom, Trenins {\em et al.}\cite{tren+19jcp} suggest calculating the quasi-centroid potential of mean force \lq\lq on the fly" during the course of a path integral molecular dynamics (PIMD) simulation, using a modified version of the adiabatic CMD algorithm.\cite{cao-voth94djcp} The trouble with this is that the adiabatic CMD algorithm is already quite expensive even with a Cartesian centroid, because of the need to use a small time step to correctly integrate the rapid oscillations of the adiabatically separated ring polymer internal modes. This problem is exacerbated in the QCMD case by the fact that the curvilinear quasi-centroid is not orthogonal to the internal modes, which complicates the algorithm.\cite{tren+19jcp} As a result, Trenins {\em et al.} found that their adiabatic QCMD calculations were 8 times more expensive than (partially) adiabatic CMD calculations for liquid water at 300~K, and 32 times more expensive for ice at 150~K.\cite{tren+19jcp}

In the case of a Cartesian centroid, there are cheaper ways to do CMD calculations. In particular, Hone {\em et al.}\cite{hone+05jcp} have suggested a \lq\lq fast CMD" method, in which the centroid forces obtained from a short PIMD simulation are least-squares fit to a pairwise model for the deviation between the classical interaction potential and the centroid potential of mean force. We have considered adapting their method to the quasi-centroid case, but found that trying to extract the quasi-centroid forces from a PIMD simulation is subject to the same difficulties that Trenins {\em et al.}\cite{tren+19jcp} encountered when developing their adiabatic QCMD algorithm. We have therefore developed a method that avoids the need to do this and focusses instead on quasi-centroid distribution functions. These distribution functions are straightforward to extract from a PIMD simulation, and they can be inverted using well-established iterative Boltzmann inversion\cite{sope96cp,reit+03jcc} methodology to give an approximation to the quasi-centroid potential of mean force that is analogous to the approximation made in fast CMD.\cite{hone+05jcp} We describe our method in Sec.~II, present some preliminary applications to gas phase molecules in Sec.~III, and discuss the prospect of further applications in Sec.~IV.

\section{Theory}

Quasi-centroid molecular dynamics, like centroid molecular dynamics, is simply classical molecular dynamics on an effective potential: the potential of mean force experienced by the quasi-centroid of the ring polymer in an imaginary time path integral simulation. The difference between the two methods is that whereas the centroid of each atom is simply the centre-of-mass of its ring polymer beads, the quasi-centroid is defined in terms of certain radial and angular coordinates that are specific to the system under investigation. This has the advantage that it eliminates the curvature problem from QCMD vibrational spectra,\cite{tren+19jcp} at the expense of some additional complexity.

Consider, for example, a gas phase water molecule, with OH bond lengths $r_1$ and $r_2$ and HOH bond angle $\theta_{12}$. At a given configuration in a $P$-bead imaginary time path integral simulation, the quasi-centroid OH bond lengths are 
\begin{equation}
\bar{r}_i = {1\over P}\sum_{j=1}^{P} {r}_i^{(j)}
\end{equation}
and the quasi-centroid HOH bond angle is
\begin{equation}
\bar{\theta}_{12} = {1\over P}\sum_{j=1}^P \theta_{12}^{(j)},
\end{equation}
where ${r}_i^{(j)}$ and $\theta_{12}^{(j)}$ are the bond lengths and the bond angle of the $j$-th molecular bead of the ring polymer necklace. The three parameters $\bar{r}_1$, $\bar{r}_2$, and $\bar{\theta}_{12}$ define the geometry of the quasi-centroid molecule. This can be oriented with the Eckart-like frame criterion described by Trenins {\em et al.}\cite{tren+19jcp} so as to bring it into alignment with the average orientation of the beads of the ring polymer necklace. The molecular centre-of-mass of the quasi-centroid can then be set equal to the average molecular centre-of-mass of the ring polymer beads to complete the specification of the quasi-centroid atomic coordinates $\bar{\bf r}_{\rm O}$, $\bar{\bf r}_{{\rm H}_1}$, and $\bar{\bf r}_{{\rm H}_2}$ of the molecule.

The situation for a gas phase ammonia molecule is similar. Here there are three quasi-centroid NH bond lengths $\bar{r}_1$, $\bar{r}_2$, and $\bar{r}_3$ and three quasi-centroid HNH bond angles $\bar{\theta}_{12}$, $\bar{\theta}_{23}$, and $\bar{\theta}_{31}$, which again suffice to define the geometry of the quasi-centroid molecule. However, $3N-6$ is only equal to $N(N-1)/2$ when the number of atoms $N$ is equal to 3 or 4. For methane, there are four CH bond lengths and six HCH bond angles, whereas only nine geometric parameters are needed to specify the shape of the quasi-centroid molecule. The problem is therefore over-determined, and the quasi-centroid coordinates can only be obtained by minimising some measure of the error in the target set of quasi-centroid bond lengths and bond angles. 

Fortunately, we can avoid this complexity when our goal is simply to construct a separable approximation to the difference between the quasi-centroid potential of mean force\cite{QCPMF} and the underlying classical interaction potential. All we require for this are the {\em distribution functions} of the quasi-centroid geometric parameters (bond lengths and bond angles), which are straightforward to extract from a PIMD simulation. Taking again the example of a gas phase water molecule, the radial distribution function of the quasi-centroid OH bond lengths is
\begin{equation}
g_r(\bar{r}_i) = {1\over 4\pi \bar{r}_i^2\rho}\left<\delta\left(\bar{r}_i-{1\over P}\sum_{j=1}^P r_i^{(j)}\right)\right>_{\rm PIMD},
\end{equation}
where $\rho$ is a constant with the dimensions of a number density and either $i=1$ or 2 can be used since since the distributions of both bond lengths are the same. The angular distribution function of quasi-centroid HOH bond angles is
\begin{equation}
g_{\theta}(\bar{\theta}_{12}) = {1\over \sin\bar{\theta}_{12}}\left<\delta\left(\bar{\theta}_{12}-{1\over P}\sum_{j=1}^P \theta_{12}^{(j)}\right)\right>_{\rm PIMD}.
\end{equation}
In both cases the angular brackets denote averages over the configurations of the ring polymer beads visited in a standard (unconstrained) PIMD simulation. The distribution functions $g_r(\bar{r}_i)$ and $g_{\theta}(\bar{\theta}_{12})$ are straightforward to calculate by accumulating histograms during this simulation. 

The reason why these distribution functions are useful is that they can also be written as classical NVT averages 
\begin{equation}
g_r(\bar{r}_i) = {1\over 4\pi \bar{r}_{i}^2\rho}\Bigl<\delta\left(\bar{r}_{i}-r_{i}\right)\Bigr>_{\rm NVT},
\end{equation}
and
\begin{equation}
g_{\theta}(\bar{\theta}_{12}) = {1\over \sin\bar{\theta}_{12}}\Bigl<\delta\left(\bar{\theta}_{12}-{\theta}_{12}\right)\Bigr>_{\rm NVT},
\end{equation}
on the quasi-centroid potential of mean force.\cite{QCPMF} Indeed, if the angular brackets in Eqs.~(5) and~(6) are interpreted as canonical (Boltzmann) averages in a classical system with the potential $V_{\rm qc}(r_1,r_2,\theta_{12})$, then it is straightforward to show that
\begin{equation}
{1\over\beta}{{\rm d}\ln g_r(\bar{r}_i)\over {\rm d}\bar{r}_i} = \left<-{\partial V_{\rm qc}(r_1,r_2,\theta_{12})\over \partial r_i}\right>_{r_i=\bar{r}_i},
\end{equation}
and
\begin{equation}
{1\over\beta}{{\rm d}\ln g_{\theta}(\bar{\theta}_{12})\over {\rm d}\bar{\theta}_{12}} = \left<-{\partial V_{\rm qc}(r_1,r_2,\theta_{12})\over \partial \theta_{12}}\right>_{\theta_{12}=\bar{\theta}_{12}},
\end{equation}
where the angular brackets denote constrained canonical averages with the indicated constraints.

It follows that we can use the distribution functions $g_r(\bar{r}_i)$ and $g_{\theta}(\bar{\theta}_{12})$ obtained from a PIMD simulation to construct a simple approximation to the quasi-centroid potential of mean force involving a separable correction to the classical interaction potential $V(r_1,r_2,\theta_{12})$,
\begin{equation}
V_{\rm qc} \simeq V+\Delta V_r(r_1)+\Delta V_r(r_2)+\Delta V_{\theta}(\theta_{12}).
\end{equation}
There are various ways to do this,\cite{sope96cp,reit+03jcc,lyub-laak95pre} the simplest of which is to use iterative Boltzmann inversion\cite{sope96cp,reit+03jcc} to invert the distribution functions $g_r({r}_i)$ and $g_{\theta}({\theta}_{12})$ and obtain the corrections $\Delta V_r(r_i)$ and $\Delta V_{\theta}(\theta_{12})$. In this method, one starts by setting $\Delta V_r^{(0)}(r_i)=0$ and $\Delta V_{\theta}^{(0)}(\theta_{12})=0$, and then iterates the equations
\begin{equation}
\Delta V_r^{(k+1)}(r_i) = \Delta V_r^{(k)}(r_i)-{1\over\beta}\ln\left[{g_r(r_i)\over g_r^{(k)}(r_i)}\right],
\end{equation} 
\begin{equation}
\Delta V_{\theta}^{(k+1)}(\theta_{12}) = \Delta V_{\theta}^{(k)}(\theta_{12})-{1\over\beta}\ln\left[{g_{\theta}(\theta_{12})\over g_{\theta}^{(k)}(\theta_{12})}\right],
\end{equation} 
to convergence, where $g_r(r_i)$ and $g_{\theta}(\theta_{12})$ are the target PIMD distribution functions and  $g_r^{(k)}(r_i)$ and $g_{\theta}^{(k)}(\theta_{12})$ are the classical distribution functions calculated using $\Delta V_r^{(k)}(r_i)$ and $\Delta V^{(k)}_{\theta}(\theta_{12})$. We used this method for all three of the molecules considered in Sec.~III and found that it converged within four iterations in every case. Examples of the convergence of $g_r^{(k)}(r_i)$ and $g_{\theta}^{(k)}(\theta_{12})$ are given in the supplementary material, along with some comments on why the iterative Boltzmann inversion converges so quickly.

Although we have focussed here on a water molecule, we should stress that exactly the same procedure can be used for ammonia and methane. There is again just one unique radial distribution function $g_r(r_i)$ and one unique angular distribution function $g_{\theta}(\theta_{ij})$ in each of these molecules by symmetry. The only new twist is that there are more radial and angular coordinates, so we have to use a more general expression for the separable correction to the quasi-centroid potential of mean force. This is
\begin{equation}
V_{\rm qc} \simeq V+\sum_{i=1}^{N_{\rm H}} \Delta V_r(r_i)+\sum_{i=1}^{N_{\rm H}}\sum_{j>i}^{N_{\rm H}} \Delta V_{\theta}(\theta_{ij}),
\end{equation}
where $N_{\rm H}=2$ for water, 3 for ammonia, and 4 for methane. Everything else carries over unaltered. (It might be interesting in future work to consider the deuterated isotopologues of the three molecules. These will have more quasi-centroid distribution functions because the mass of a particle affects its quantum mechanical dispersion. We have not yet done this but we can see no reason why the present method should not be able to cope with either partially or fully deuterated molecules.)

\section{Results and Discussion}

\begin{figure*}[t!]
\centering
\resizebox{1.8\columnwidth}{!} {\includegraphics{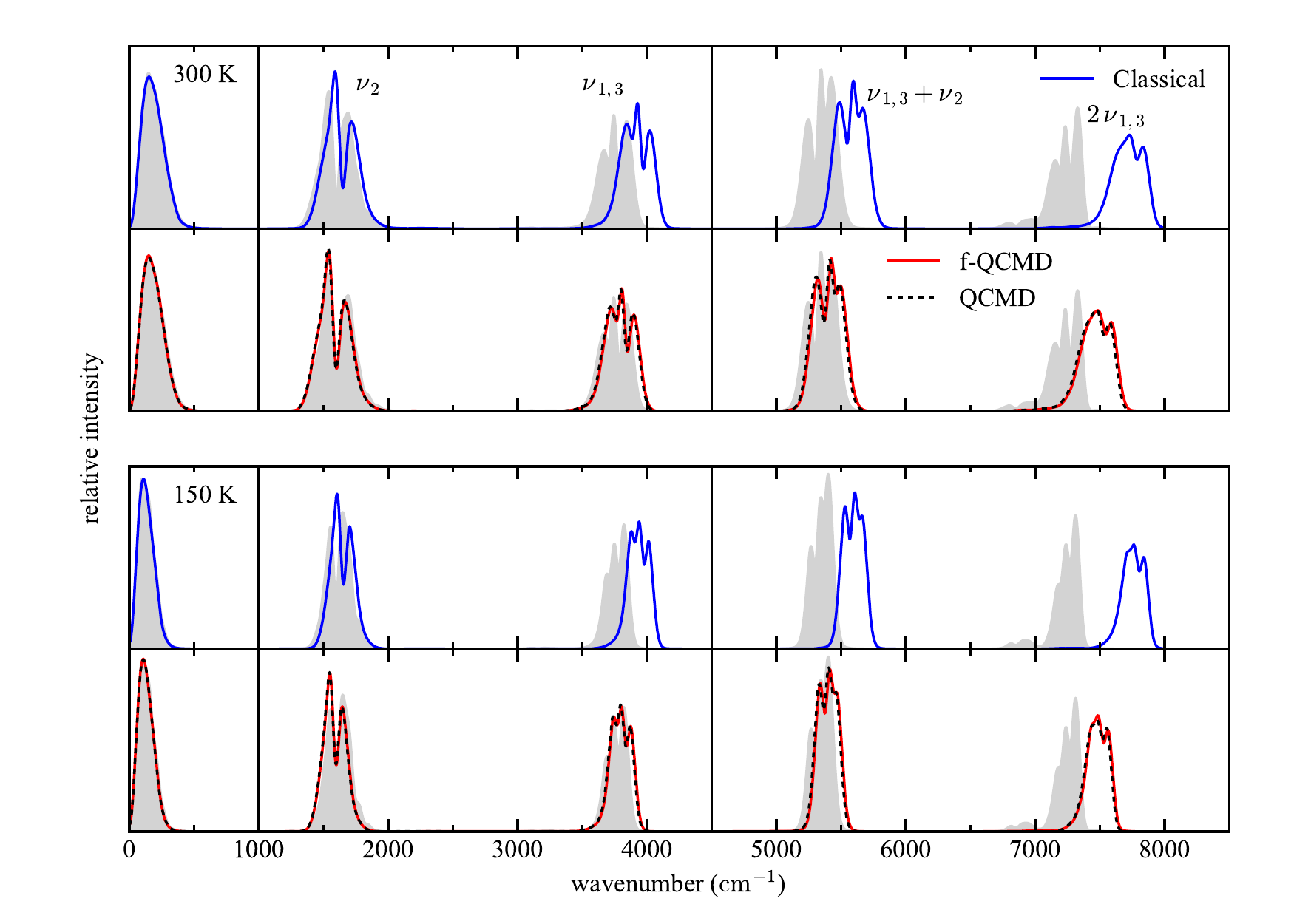}}
\caption{Quantum (shaded), classical (blue), fast QCMD (red), and adiabatic QCMD (dashed) vibrational spectra of a water molecule at 300~K (top) and 150~K (bottom). The relative absorption intensities in the three panels are scaled in the ratio 1:5:70. The adiabatic QCMD results and the quantum mechanical reference spectra were provided by the authors of Ref.~\onlinecite{bens+20fd}.}
 \end{figure*}

\subsection{Water}

Our fast quasi-centroid molecular dynamics (f-QCMD) vibrational spectra of a water molecule at 300~K and 150~K are compared with the quantum mechanical, classical, and adiabatic QCMD spectra in Fig.~1. The resolutions of all spectra are the same, corresponding to convolution with the Fourier transform of a Hann window function $f(t) = \cos^2(\pi t/2T)$ with $T=0.75$ ps. The classical spectra were calculated as Fourier transforms of classical dipole time derivative autocorrelation functions. The intensities of the combination and overtone bands were then corrected using the formula of Yao and Overend,\cite{yao-over76saa} as discussed in the recent papers by Benson and Althorpe\cite{bens-alth21jcp} and by Pl\'e {\em et al.}\cite{ple+21jcp} The f-QCMD spectra were calculated in the same way, using the quasi-centroid potentials of mean force obtained from iterative Boltzmann inversion of quasi-centroid distribution functions generated in PIMD simulations with $P=64$ ring polymer beads. The quantum mechanical and adiabatic QCMD spectra were obtained from Benson {\em et al.},\cite{bens+20fd} who calculated the former using the DVR3D program of Tennyson and co-workers.\cite{tenn+04cpc} All spectra were calculated using the Partridge-Schwenke potential energy surface\cite{part-schw97jcp} and dipole moment function.\cite{schw-part00jcp}
 
 The results in Fig.~1 show that the f-QCMD approximation does a good job of correcting the vibrational band frequencies  of the purely classical simulation and bringing them into closer agreement with those of the quantum calculation. There is no artificial broadening of the band profiles as would be seen in TRPMD, or any artificial red shifting of the stretching band at 150~K as would be seen in CMD.\cite{bens+20fd} The f-QCMD bending band ($\nu_2$) is in excellent agreement with the quantum calculation. The f-QCMD stretching band ($\nu_{1,3}$) is still slightly blue-shifted relative to the quantum calculation, but not by nearly so much as in the classical case. The blue shift is also present in the $\nu_{1,3}+\nu_2$ combination band, and it is exaggerated in the $2\nu_{1,3}$ overtone band. Identical blue shifts are seen in the adiabatic QCMD reference calculations of Benson {\em et al.},\cite{bens+20fd} and indeed our f-QCMD results are in excellent agreement with those of these reference calculations throughout the frequency range shown in Fig.~1. The separable approximation to the quasi-centroid potential of mean force in Eq.~(12) is therefore perfectly adequate for a gas phase water molecule.
 
\begin{figure*}[t]
\centering
\resizebox{1.8\columnwidth}{!} {\includegraphics{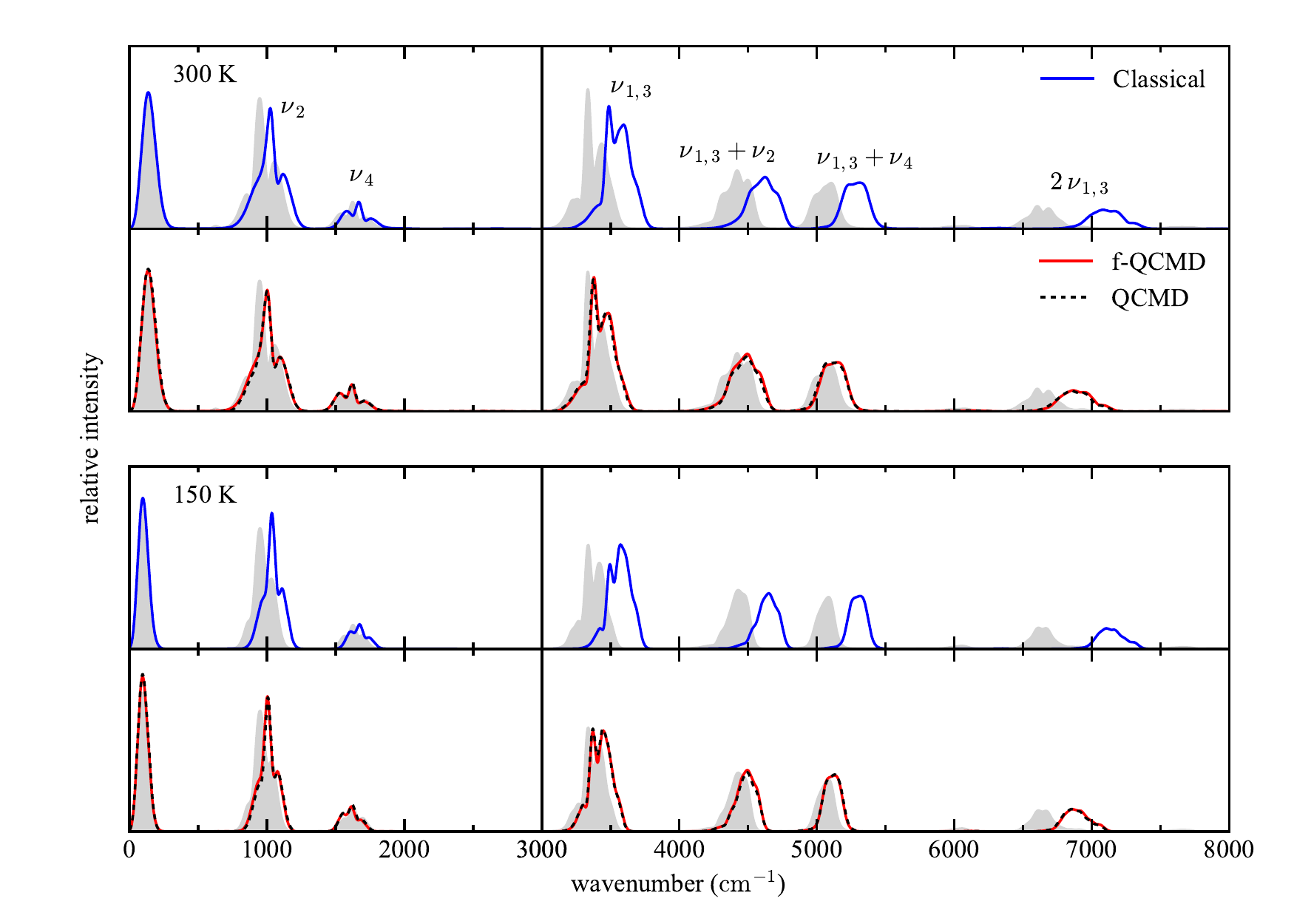}}
\caption{Quantum (shaded), classical (blue), fast QCMD (red), and adiabatic QCMD (dashed) vibrational spectra of an ammonia molecule at 300~K (top) and 150~K (bottom). The relative absorption intensities in the two panels are scaled in the ratio 1:14. The adiabatic QCMD results were provided by the authors of Ref.~\onlinecite{hagg+21jcp}.}
 \end{figure*}
 
\subsection{Ammonia}

Our f-QCMD results for ammonia are compared with the quantum, classical, and adiabatic QCMD results in Fig.~2. The classical and f-QCMD calculations were performed in the same way as for water, using the PES-2 potential energy surface and the AQZfc dipole moment surface of Yurchenko {\em et al.}\cite{yurc+09jpca} The quantum mechanical spectra were extracted from the variationally computed line list given in the Supplementary Information of Ref.~\onlinecite{yurc+09jpca}, and the adiabatic QCMD spectra were obtained from Ref.~\onlinecite{hagg+21jcp}.

The f-QCMD results in Fig.~2 are again seen to do a good job of correcting the classical band frequencies without broadening the band profiles. The fundamental region of the f-QCMD spectrum is in good agreement with the quantum spectrum apart from slight blue shifts in the $\nu_{1,3}$ stretching and $\nu_2$ bending bands. This is again mirrored in the combination bands involving these fundamentals and exaggerated in the $2\nu_{1,3}$ overtone band, just as was seen for water in Fig.~1. The QCMD reference spectra in Fig.~2 were calculated by Haggard {\em et al.}\cite{hagg+21jcp} using the adiabatic QCMD algorithm. The present f-QCMD results are in excellent agreement with these reference spectra throughout the frequency range shown in the figure. This confirms that the separable approximation to the centroid potential of mean force in Eq.~(12) is just as accurate for ammonia as it is for water.

Since the QCMD and f-QCMD spectra for ammonia are the same, we shall defer to the paper of Haggard {\em et al.}\cite{hagg+21jcp} for a fuller discussion, including in particular an explanation for the blue shift in the QCMD $\nu_2$ bending band relative to the quantum mechanical calculation. This is associated with the formation of instanton tunnelling paths along the $\nu_2$ coordinate. These tunnelling paths lead to ammonia inversion events in the quantum calculation that are suppressed in the QCMD calculation, resulting in a higher $\nu_2$ bending frequency.\cite{hagg+21jcp}  

\begin{figure*}[t]
\centering
\resizebox{1.8\columnwidth}{!} {\includegraphics{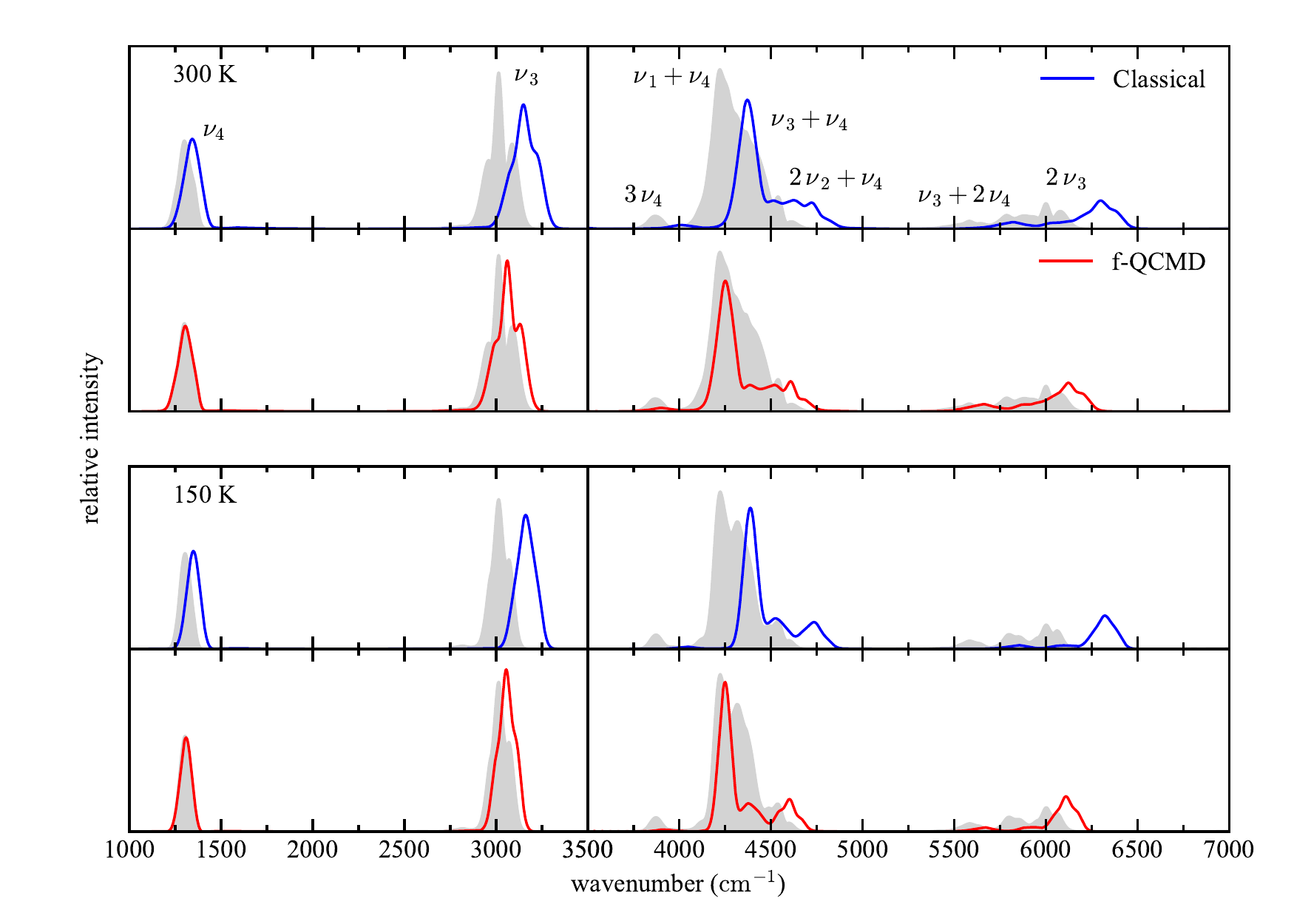}}
\caption{Quantum (shaded), classical (blue), and fast QCMD (red) vibrational spectra of a methane molecule at 300~K (top) and 150~K (bottom). The relative absorption intensities in the two panels are scaled in the ratio 1:28. (Methane does not have a permanent dipole moment so the rotational band below 1000 cm$^{-1}$ has negligible intensity.) In this case there are no adiabatic QCMD reference spectra available for comparison.}
 \end{figure*}

\subsection{Methane}

Our last example is an application of f-QCMD to the vibrational spectrum of methane. This is a more challenging problem because the quasi-centroid of the methane molecule is over-determined by the full set of quasi-centroid CH bond lengths and HCH bond angles, as we have discussed in Sec.~II. This is not an issue for f-QCMD, but it would make an adiabatic QCMD calculation more difficult, which may explain why there is as yet no adiabatic QCMD reference spectrum for methane available for comparison.  
 
Our classical and f-QCMD calculations for methane were performed in the same way as our calculations for water and ammonia, using the {\em ab initio} potential energy and dipole moment surfaces of Yurchenko {\em et al.}\cite{yurc+13jms} The quantum mechanical reference spectra were extracted from the ExoMol line list reported in Ref.~\onlinecite{yurc-tenn14mnras}, which was calculated using the same potential energy and dipole moment surfaces.

The results of these methane calculations are shown in Fig.~3. In this case the f-QCMD spectrum is again a significant improvement on the classical spectrum. The f-QCMD calculation reproduces the quantum mechanical IR-active $\nu_4$ bending band almost exactly, it gives a $\nu_3$ stretching band that is only slightly blue shifted from the quantum spectrum, and it shifts the classical frequencies of the overtone and combination bands into much better agreement with the quantum mechanical calculation. If this level of agreement between f-QCMD and quantum mechanics continues to hold for larger molecules then f-QCMD could become quite a useful tool for predicting low resolution vibrational spectra. 

There is however an issue with the intensities of the f-QCMD combination and overtone bands in Fig.~3, especially in the region of the $\nu_1+\nu_4/\nu_3+\nu_4/2\nu_2+\nu_4$ combinations. This issue arises because, not knowing the f-QCMD frequencies of the IR-inactive $\nu_1$ and $\nu_2$ fundamentals, we have simply used the two-mode $\nu_3+\nu_4$ combination band formula of Yao and Overend\cite{yao-over76saa} to adjust the computed intensity in this region of the spectrum. Similar remarks apply to the intensity in the region of the $\nu_3+2\nu_4$ combination and $2\nu_3$ overtone bands, for which we have simply used the two-mode $2\nu_3$ overtone band formula. It is certainly possible that one could do better than this, for example by calculating a Raman spectrum to reveal the f-QCMD frequencies of the IR-inactive fundamentals and by using appropriate intensity correction factors for classical 3-mode combination bands. However, we suspect that the intensities of overlapping combination and overtone bands, which are ubiquitous in the spectroscopy of methane,\cite{ulen+14jcp} will always be a challenge for trajectory-based methods such as f-QCMD.

 \section{Conclusions and future work}
 
In this paper, we have described a fast implementation of the QCMD method, verified that it gives good agreement with adiabatic QCMD calculations for the vibrational spectra of gas phase water and ammonia, and shown that it also gives good agreement with the quantum mechanical IR spectrum of methane.

The only path integral component of f-QCMD is a short PIMD simulation, which is used to calculate quasi-centroid radial and angular distribution functions. The remaining two components of the method, the iterative Boltzmann inversion of the distribution functions to obtain an approximation to the quasi-centroid potential of mean force and the subsequent calculation of the vibrational spectrum on this potential of mean force, only involve classical molecular dynamics simulations. The numerical effort of the method is therefore minimal, and should not present any obstacle to applying it to more complex systems.

We have already begun to investigate the application of f-QCMD to the spectrum of the CH$_5^+$ molecular ion,\cite{whit+99science,brow+jcp04,asva+05science} which is a significantly more challenging and interesting system than the methane molecule we have considered in Fig.~3. Once that has been done, the next stage will be to test the method for a molecule with torsional vibrations. With torsions included, it should be possible to apply the method to any molecule (or biomolecule) for which appropriate potential energy and dipole moment surfaces are available. (The availability of polarisability surfaces would also allow for the calculation of Raman spectra.) There is also no reason why the method could not be applied to molecular liquids and solids, by augmenting the intramolecular quasi-centroid distribution functions we have discussed here with intermolecular quasi-centroid radial distribution functions. The resulting quasi-centroid potentials of mean force could be used not only for the calculation of vibrational spectra, but also for the calculation of QCMD transport coefficients. For example the thermal conductivity could be calculated on the quasi-centroid potential of mean force using the method described in Ref.~\onlinecite{suth+21jcp}.

One final comment is that our results for water, ammonia, and methane can be used to shed light on the errors to be expected in the frequencies of QCMD combination and overtone bands. For all three molecules, the method does a good job of capturing the anharmonic red shifts in the fundamental bands that are missed in the classical simulation. However, the QCMD combination and overtone bands are peaked at frequencies close to the sum of the frequencies of the contributing fundamentals, rather than exhibiting any further anharmonic red shifts that might be present in an exact quantum calculation. In fact, a simple argument based on Morse oscillator energy levels suggests that the error in the frequency of a QCMD combination (difference) band will be the sum (difference) of the errors in the contributing fundamentals, and that the error in the frequency of a QCMD overtone will be three times the error in the QCMD fundamental plus the difference between the frequencies of the classical and QCMD fundamentals.\cite{Morse} These errors are consistent with the results in our figures and they should be borne in mind in future applications of f-QCMD to more complex systems. \\

\section*{Supplementary Material}
The supplementary material illustrates the convergence of the iterative Boltzmann inversion in Eqs.~(10) and~(11) and comments on why the iteration converges so quickly.

\begin{acknowledgements} 
We are grateful to Stuart Althorpe, Raz Benson, Christopher Haggard, and Vijay Ganesh Sadhasivam for helping us to validate our f-QCMD calculations by comparison with their adiabatic QCMD calculations, and for several interesting discussions. Theo Fletcher is supported by the EPSRC Centre for Doctoral Training in Theory and Modelling in the Chemical Sciences, EPSRC Grant No. EP/L015722/1, and Joseph Lawrence is supported by an ETH Z\"urich Postdoctoral Fellowship.
\end{acknowledgements}

\section*{Data Availability}

The data that support the findings of this study are available within the paper and the supplementary material.


\begin{thebibliography}{99}
\bibitem{Morse}
It follows that a significant part of the error in the frequency of a QCMD overtone can be eliminated by subtracting the difference between the frequencies of the classical and QCMD fundamentals, both of which are readily available without doing an exact quantum mechanical calculation.
\end{thebibliography}

\end{document}


\title{Fast quasi-centroid molecular dynamics}
\author{Theo Fletcher}
\email{theo.fletcher@chem.ox.ac.uk}
\affiliation{Department of Chemistry, University of Oxford, Physical and Theoretical Chemistry Laboratory, South Parks Road, Oxford, OX1 3QZ, UK}
\author{Andrew Zhu}
\affiliation{Department of Chemistry, University of Oxford, Physical and Theoretical Chemistry Laboratory, South Parks Road, Oxford, OX1 3QZ, UK}
\author{Joseph E. Lawrence}
\affiliation{Laboratory of Physical Chemistry, ETH Z\"urich, 8093 Z\"urich, Switzerland}
\author{David E. Manolopoulos}
\affiliation{Department of Chemistry, University of Oxford, Physical and Theoretical Chemistry Laboratory, South Parks Road, Oxford, OX1 3QZ, UK}

\renewcommand\thefigure{\Alph{figure}} 
\maketitle

\section{Supplementary Material}

The following figures illustrate the convergence of the iterative Boltzmann inversion (IBI) for ammonia, at 300 K in Fig.~A and at 150 K in Fig.~B. Decreasing the temperature does not seem to affect the number of iterations required for convergence. The convergence for water and methane was found to be faster, with full convergence reached after 3 iterations for water and just 2 iterations for methane (at both temperatures). We believe the reason why the IBI converges so quickly in the present context is that we are not asking too much of it: we are simply using it to calculate a small correction to the classical interaction potential rather than the full quasi-centroid potential of mean force. 
 
\begin{figure}[h!]
    \centering
    \includegraphics[width=\linewidth]{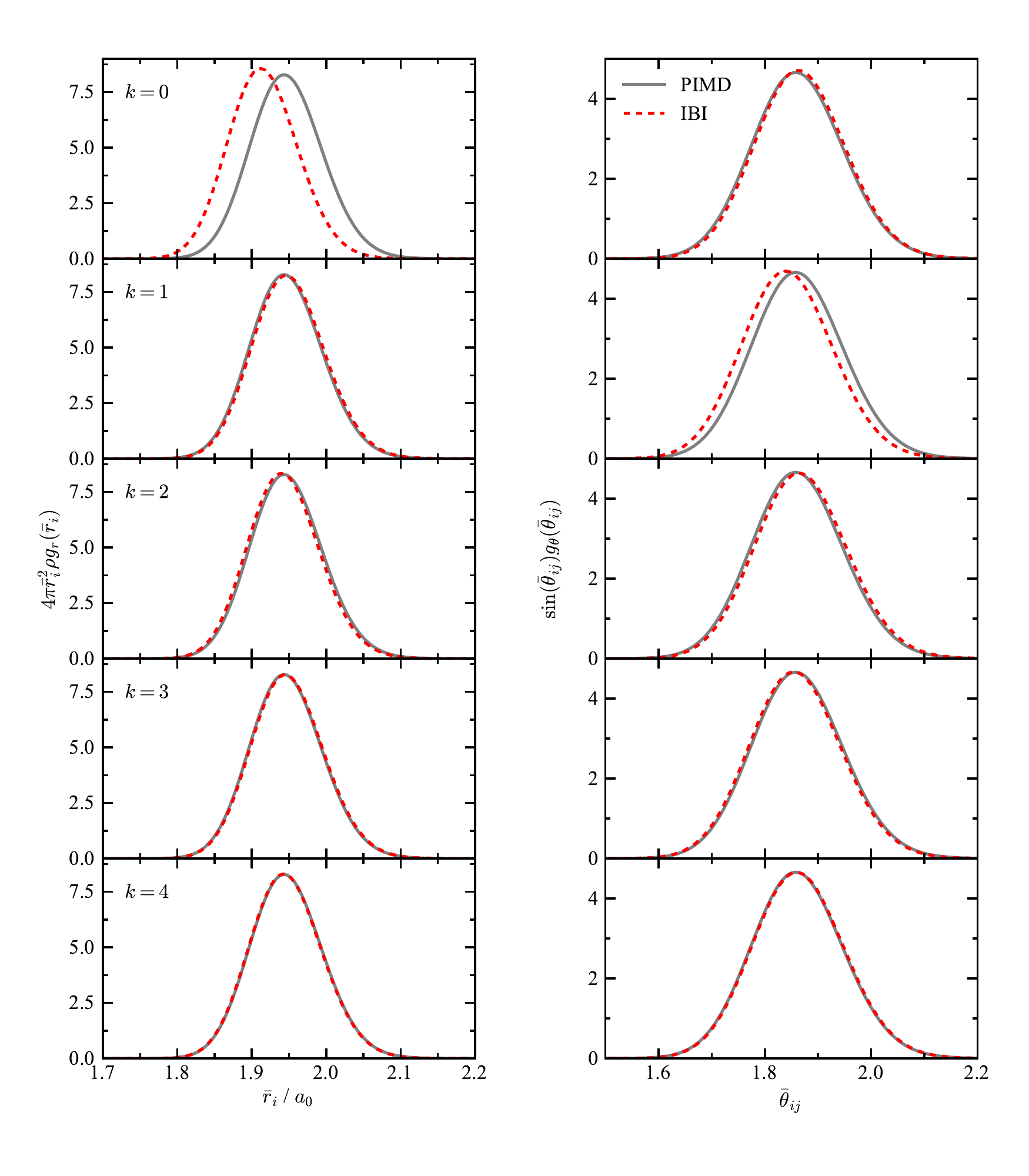}
    \caption{Quasi-centroid NH bond length and HNH bond angle distributions of ammonia at 300 K (solid grey lines), and at each stage of the iterative Boltzmann inversion (dashed red lines). The dashed distributions at iteration $k=0$ are the purely classical distributions that are used as a starting point.}
    \label{fig:ibi_convergence}
\end{figure}
 
\begin{figure}[h!]
    \centering
    \includegraphics[width=\linewidth]{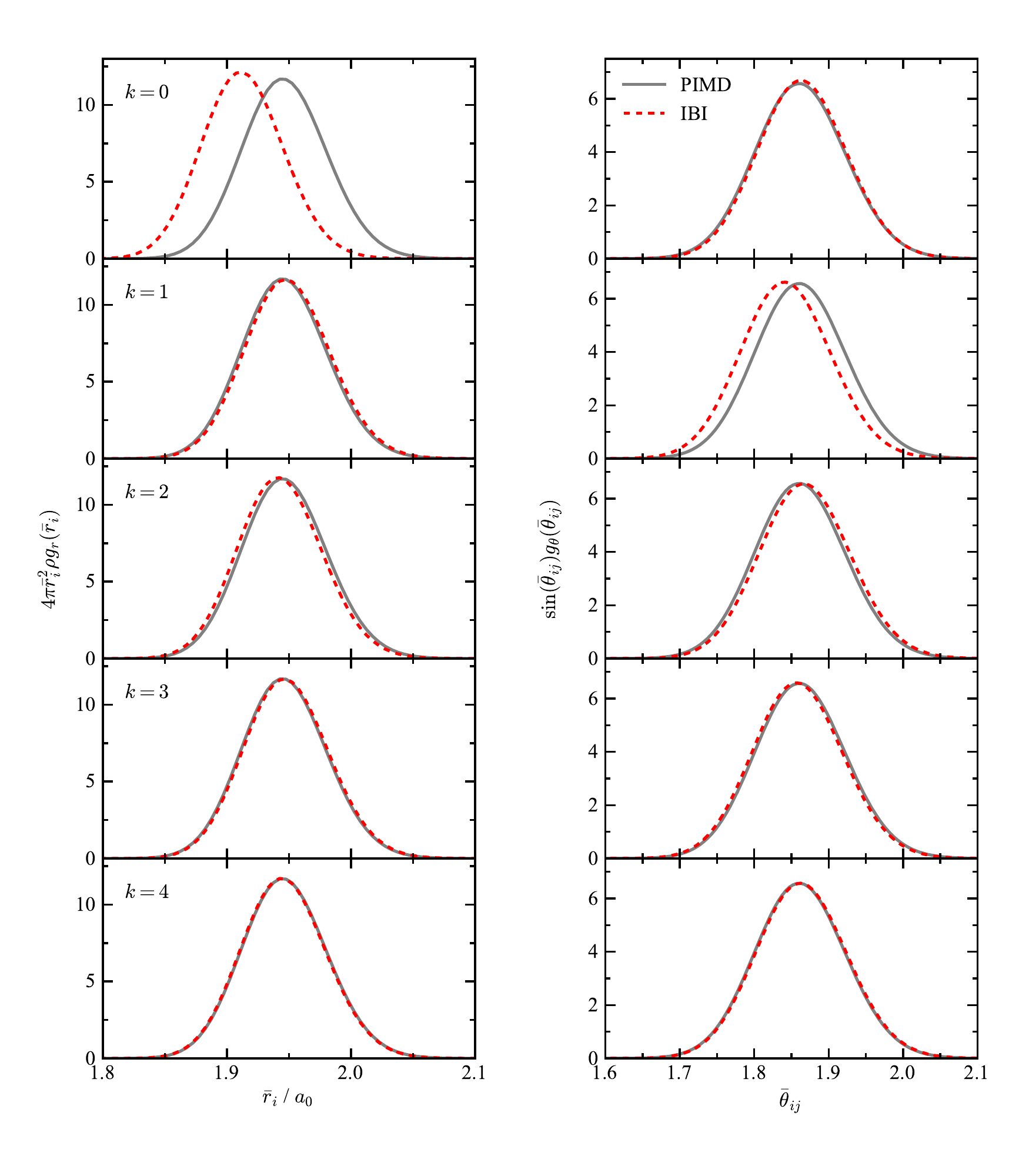}
    \caption{Quasi-centroid NH bond length and HNH bond angle distributions of ammonia at 150 K (solid grey lines), and at each stage of the iterative Boltzmann inversion (dashed red lines). The dashed distributions at iteration $k=0$ are the purely classical distributions that are used as a starting point.}
    \label{fig:ibi_convergence}
\end{figure}